\documentclass[11pt,a4paper]{article}
\pdfoutput=1
\usepackage[left=1in,right=1in,top=1in,bottom=1in]{geometry}
\usepackage[T1]{fontenc}
\usepackage{graphicx}
\graphicspath{{./Figures/}}
\usepackage{float}
\usepackage{subcaption}
\usepackage[tbtags]{amsmath}
\usepackage{amssymb}
\usepackage{amsthm}
\usepackage{hyperref}
\usepackage[numbers]{natbib}
\newtheorem{lemma}{Lemma}
\def \R{\mathbb R}
\newcommand{\sset}[1]{\left\{ #1\right\}}
\newcommand{\map}{\longrightarrow}
\newcommand{\algoname}[1]{\ensuremath{\text{\rm\sc #1}}}
\newcommand{\vecc}[1]{\ensuremath{\mathbf{#1}}} 

\title{A Note on Selling Optimally Two Uniformly Distributed Goods\footnote{This work was supported by ERC Advanced Grant 321171 (ALGAME) and the European Union Seventh Framework Programme (FP7/2007-2013) grant 284731 (UaESMC).}}
\author{Yiannis Giannakopoulos\thanks{Department of Computer Science, University of Oxford. Email: \href{mailto:ygiannak@cs.ox.ac.uk}{\nolinkurl{ygiannak@cs.ox.ac.uk} }}}
\date{February 8, 2015}
\begin{document}
\maketitle

\begin{abstract}
We provide a new, much simplified and straightforward proof to a result of \citet{Pavlov:2011fk} regarding the revenue maximizing mechanism for selling two goods with uniformly i.i.d.~valuations over intervals $[c,c+1]$, to an additive buyer. This is done by explicitly defining optimal dual solutions to a relaxed version of the problem, where the convexity requirement for the bidder's utility has been dropped. Their optimality comes directly from their structure, through the use of exact complementarity. For $c=0$ and $c\geq 0.092$ it turns out that the corresponding optimal primal solution is a feasible selling mechanism, thus the initial relaxation comes without a loss, and revenue maximality follows. However, for $0<c<0.092$ that's not the case, providing the first clear example where relaxing convexity provably does not come for free, even in a two-item regularly i.i.d.~setting. 
\end{abstract}

\section{Introduction}
In this short note we deal with the problem of maximizing the expected revenue of a two-good monopolist when facing an additive buyer whose values for the goods come uniformly i.i.d.~over intervals $[c,c+1]$, $c>0$. Notice that the optimization here is not only over deterministic mechanisms (i.e. price schedules), but all possible lotteries as well. The general model of such additive bayesian auctions has been extensively studied in the last years and the particular problem was solved by~\citet{Pavlov:2011fk}. In the case of $c=0$, the optimal selling mechanism is deterministic with prices $2/3$ for each of the items and $(4-\sqrt{2})/3$ for their bundle. This result was already known by the work of~\citet{Manelli:2006vn}, and an alternative proof based on duality and complementarity can be found also in~\citep{gk2014}. For $c\geq 0.077$, the optimal mechanism is again deterministic and it only offers the full bundle for a price of $(4c+\sqrt{4c^2+6})/3$. For the range in between, that is for $c\in (0,0.077)$, \citeauthor{Pavlov:2011fk} numerically computes that the optimal mechanism is a randomized one, with a menu-complexity~\citep{Hart:2012ys} of $4$. 

Here, we present a very simple alternative proof for the cases of $c=0$ and $c\geq 0.092$. For the remaining case, although we give the optimal solution to the primal-dual framework upon which our proof technique is based, it turns out that it is \emph{not} convex and that its objective value is strictly greater than the optimal revenue that can be achieved by any feasible selling mechanism. This is because our primal program is a relaxed version of the original revenue-maximization one, dropping the convexity constraint for the bidder's utility function. This relaxation is the standard approach so far in duality theory frameworks for such problems (see~\cite{Daskalakis:2013vn,gk2014}). So, \emph{this is a demonstration of the necessity, in general, of the convexity requirement for exact optimal mechanism design, even in the case of two regularly i.i.d.~items\footnote{An example of the necessity of convexity is also given in~\cite[Appendix~C]{gk2014}, even for one item, but the distribution used there is not regular.}.} Nevertheless, on the positive side, we are able to demonstrate that the two solutions are practically very close to each-other (within a factor of 7.5\textpertenthousand): in Fig.~\ref{fig:ratios_lp_auction_small_a} we provide upper bounds on approximation ratios for that optimal relaxed primal value with respect to the revenue achieved by the best randomized, the best deterministic and the best full-bundling mechanisms.  

A characteristic of our method that differentiates it from previous revenue maximization results that use duality frameworks is that it is completely constructive: \emph{we give explicit, simple closed-form definitions of the dual functions, rather than just proving their existence.} This immediately allows for a trivial check of optimality: just compute their (dual objective) value and see if this coincides with the revenue induced by the (primal) utility function. However, we follow an even simpler way: we deploy (tight) complementarity from the framework of \citet{gk2014} and so we can verify their optimality just by looking at some simple features of their structure.

\section{Primal-Dual Formulation}
The problem of maximizing revenue in our setting boils down to maximizing
\begin{equation}
\label{eq:revenue}
\int_c^{c+1}\int_c^{c+1}\frac{\partial u(\vecc x)}{\partial x_1}x_1+\frac{\partial u(\vecc x)}{\partial x_2}x_2-u(\vecc x)\,d\vecc x
\end{equation}
over the space of all convex\footnote{Convexity comes from the requirement that the selling mechanisms need to be truthful, that is, the buyer has no incentive to misreport her true valuation profile $\vecc x$ (see e.g.~\cite{Hart:2012uq}). Notice that this restriction is without loss with respect to the revenue maximization objective, due to the Revelation Principle\cite{Myerson:1981aa}.} functions $u:[c,c+1]^2\map\R_{\geq 0}$ with partial derivatives in $[0,1]$. Then, the optimal mechanism can be recovered completely by the bidder's utility function $u$ (see e.g.~\citep{Rochet:1985aa,Hart:2012uq}): the probability of allocating item $j$ to the bidder when she declares bids $\vecc x=(x_1,x_2)$ is $\partial u(\vecc x)/\partial x_j$ and the payment she has to submit equals $\nabla u(\vecc x)\cdot \vecc x-u(\vecc x)$. For example, that means that any deterministic mechanism for our setting is induced by a utility of the form $u(\vecc x)=\max\sset{0,x_1-p_1,x_2-p_2,x_1+x_2-p}$, where $p_1=p_2$ \footnote{Symmetry here comes without a loss, see e.g.~\citep[Appendix~1]{Hart:2012uq}.} is the price the seller sets for each one of the items and $p$ the price for their combined bundle. In particular, the work of \citet{Pavlov:2011fk} tells us that for $c=0$ revenue~\eqref{eq:revenue} is maximized by $u(\vecc x)=\max\sset{0,x_1-\frac{2}{3},x_2-\frac{2}{3},x_1+x_2-\frac{4-\sqrt{2}}{3}}$ and for $c\geq 0.077$ by $u(\vecc x)=\max\sset{0,x_1+x_2-(4c+\sqrt{4c^2+6})/3}$.

For ease of reference, we briefly mention below the tools we'll need from the duality framework of~\citep{gk2014}. We relax the above initial problem of revenue maximization by dropping the convexity constraint on $u$, as well as the lower bound on the derivatives $\nabla u \geq \vecc{0}$. This results in the following optimization problem which we'll refer to as \emph{primal}: $\sup_{\tilde u} \int_{[c,c+1]^2}\nabla u(\vecc x)\cdot\vecc x-u(\vecc x)$ where $\tilde u:[c,c+1]^2\map\R_{\geq 0}$ is absolutely continuous with
\begin{equation}
\label{eq:primal_const_1}
\frac{\partial u(\vecc x)}{\partial x_1},\frac{\partial u(\vecc x)}{\partial x_2} \leq 1
\end{equation}
for almost every (a.e.) $\vecc x\in [c,c+1]^2$. Its \emph{dual} is: $\inf_{z_1,z_2}\int_{[c,c+1]^2}z_1+z_2$ where $z_j:[c,c+1]^2\map\R_{\geq 0}$ is absolutely continuous with respect to its $j$-th coordinate and
\begin{align}
\frac{\partial z_1(\vecc x)}{\partial x_2}+\frac{\partial z_2(\vecc x)}{\partial x_2} &\leq 3 \label{eq:dual_const_1}\\
z_1(c,x_2), z_2(x_1,c) &\leq c, \label{eq:dual_const_2}\\
z_1(c+1,x_2), z_1(x_1,c+1) &\geq c+1 \label{eq:dual_const_3},
\end{align}
a.e. in $[c,c+1]^2$. Then, the following tool which strongly resembles traditional LP complementary slackness can be used to demonstrate tight solutions:

\begin{lemma}[Exact Complementarity]
\label{lemma:complementarity}
If for almost every $\vecc x\in [c,c+1]$ the following conditions hold for a pair of primal and dual solutions $\tilde u$ and $z_1,z_2$ then they are both optimal:
\begin{itemize}
\item Either $\tilde u(\vecc x)$ is zero or the dual constraints~\eqref{eq:dual_const_1}--\eqref{eq:dual_const_3} hold with strict equality
\item For any $j=1,2$, either $z_j(\vecc x)$ is zero or the corresponding primal constraint in~\eqref{eq:primal_const_1} holds with strict equality.
\end{itemize}
\end{lemma} 

\section{The case of $0\leq c\leq 0.092$}
\label{sec:uniform_gen_1}
\begin{figure}
        \centering
        \begin{subfigure}[t]{0.6\textwidth}
                \includegraphics[width=\textwidth]{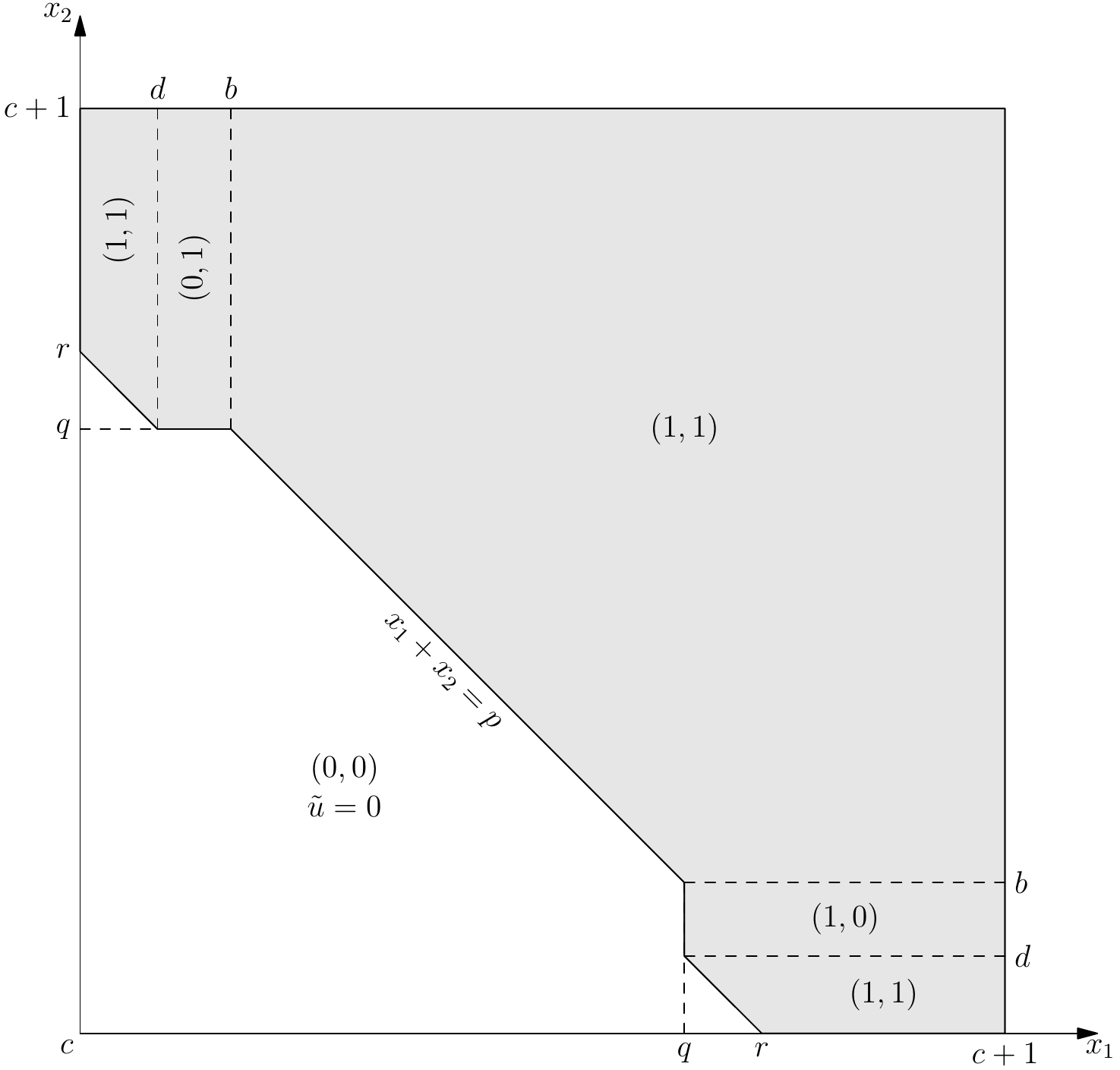}
                \caption{The values of $\nabla \tilde u(\vecc x)$ of an optimal primal solution $\tilde u$.}
                \label{fig:uniform_1_primal}
        \end{subfigure}
        ~
        \begin{subfigure}[t]{0.6\textwidth}
                \includegraphics[width=\textwidth]{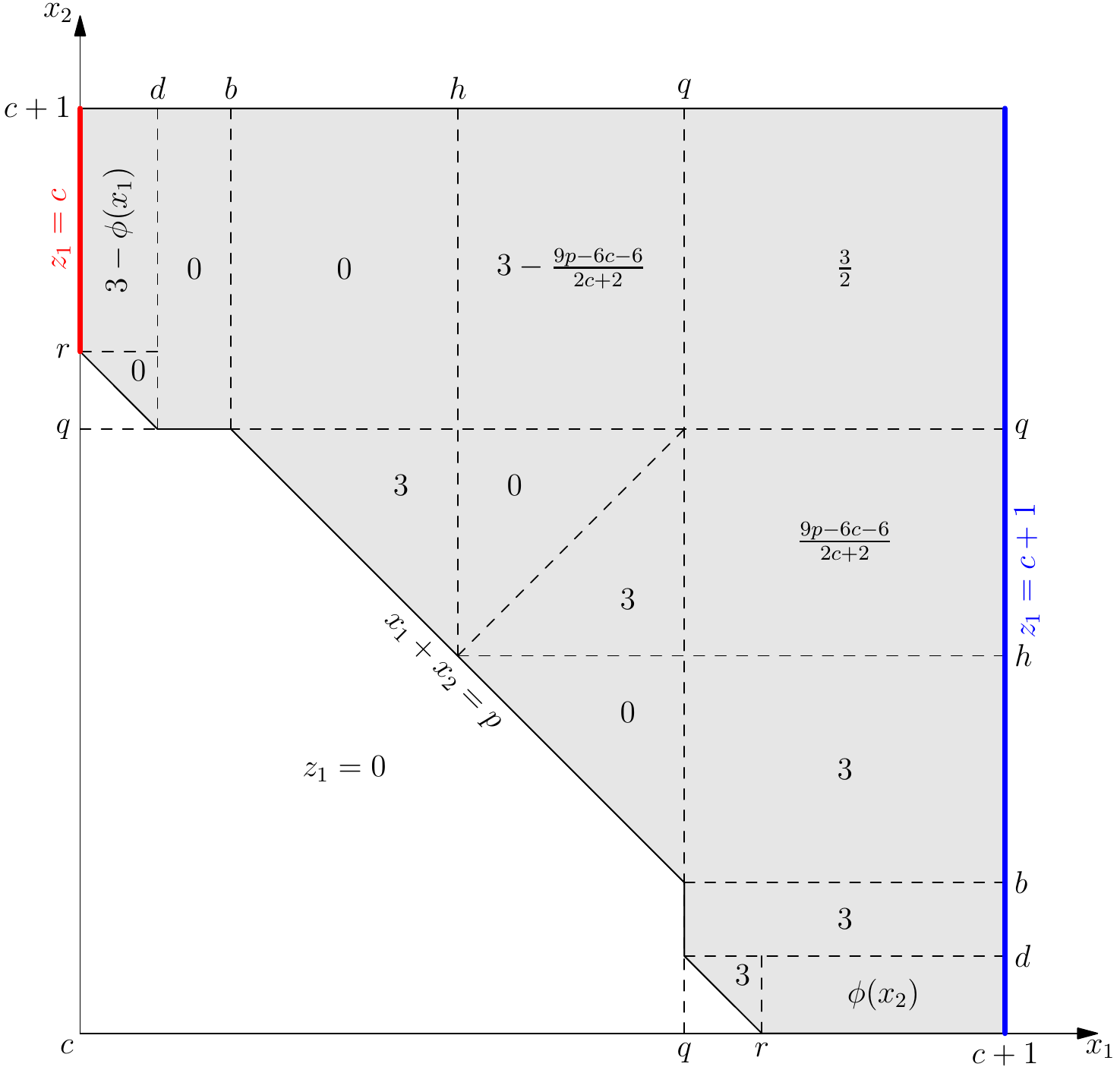}
                \caption{The values of $\partial z_1(\vecc x)/\partial x_1$. The duals are symmetric and so $\partial z_2(\vecc x)/\partial x_2$ can be recovered by the relation $\partial z_2/\partial x_2=3-\partial z_1/\partial x_1$ in the gray area (and $z_2=0$ in the white). Here $\phi(x)=\frac{c+1-3(x-c)}{c+1-r}.$}
                \label{fig:uniform_1_dual}
        \end{subfigure}
        \caption{A pair of optimal primal-dual solutions for $0\leq c\leq 0.092$. Notice how the primal solution $\tilde u$ in Fig.~\ref{fig:uniform_1_primal} is \emph{not} convex, so it does not correspond to a valid utility function of a truthful selling mechanism. The values of the various parameters are: $q=\frac{2(c+1)}{3}$, $p=\frac{4-\sqrt{2}}{3}(c+1)$, $h=\frac{p}{2}$, $b=p-q$, $r=\frac{1}{3}\left(2+c+\sqrt{c(2+3c)}\right)$, $d=\frac{1}{3}\left(2c+\sqrt{c(2+3c)}\right)$.}
        \label{fig:uniform_1}
\end{figure}
Consider the pair of primal-dual variables $\tilde u, (z_1,z_2)$ whose derivatives are given in Fig.~\ref{fig:uniform_1}. The duals $z_1$, $z_2$ are symmetric, in the sense that $z_1(x_1,x_2)=z_2(x_2,x_1)$ for all $\vecc x$. Notice that this is enough to completely define them, by the initial conditions $\tilde u(\vecc x)=0$  and $z_1(\vecc x)=0$, $z_1(\vecc x)=c$ given in the white and red areas. We will argue that they are optimal. By looking at their structure in Fig.~\ref{fig:uniform_1}, it is easy to see that the applicability of the (exact) complementarity Lemma~\ref{lemma:complementarity} is just a matter of simple calculations, essentially to check constraints~\eqref{eq:dual_const_1}--\eqref{eq:dual_const_3}. Optimality would be immediate.

To do that, first of all we need to check that the specific parameters give rise to a consistent partitioning of the allocation space, and in particular that $c\leq d\leq b\leq q\leq r\leq c+1$, $r-q=d-c$ and $p=q+b$. Given the choice of the parameters, it is trivial to check that the two last equalities are satisfied. The first chain of inequalities is satisfied for all $0\leq c\leq \bar c$ where $\bar c= \sqrt{15-8 \sqrt{2}}-2 \sqrt{2}+1\approx 0.0915$. At this value $c=\bar c$ we get the limiting situation when $d=b$ and $p=r$, and $\tilde u$ is a feasible utility function of the mechanism that offer only the full-bundle for a price of $p$. On the other hand, notice that for $c=0$ we get $q=r$ and $d=c$, and the pair of primal-duals naturally reduces to the well-know optimal selling mechanism for two uniform items on $[0,1]$ with prices $q=\frac{2}{3}$ and $p=\frac{4-\sqrt{2}}{3}$ for the one- and two-item bundles respectively.

We now just have to show that $z_1$ is feasible. In particular, it is again easy to calculate that $z_1(c+1,x_2)=c+1$, given the values of the parameters, the definition of $\phi$ and the initial condition $z_1(a,x_2)=c$ for $x_2>r$ and $z_1(c,x_2)=0$ otherwise. The only thing remaining to check is that $z_1$ never falls below zero. This can be done by easily verifying that indeed $c+\int_c^d3-\phi(t)\,dt=0$, so $z_1$ is nonnegative at the upper critical stripe. Everywhere else, all its derivatives are nonnegative, so it cannot decrease further.

\subsection{The Necessity of Convexity}
\begin{figure}
        \centering
        \begin{subfigure}[t]{0.6\textwidth}
                \includegraphics[width=\textwidth]{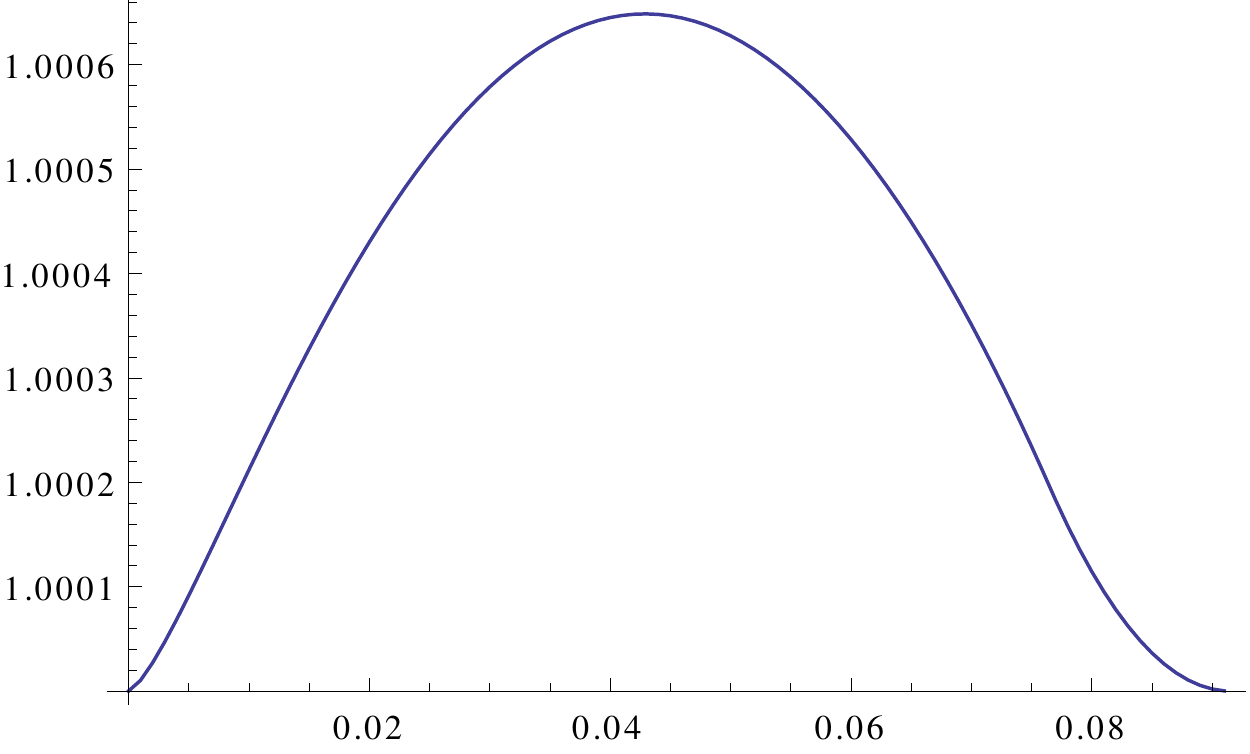}
                \caption{Best randomized selling mechanism, $1.00075$-approximate. To plot this one can use the form of the optimal auction given in~\citep{Pavlov:2011fk}.}
                \label{fig:ratios_lp_auction_rand_small_a}
        \end{subfigure}
        ~
        \begin{subfigure}[t]{0.6\textwidth}
                \includegraphics[width=\textwidth]{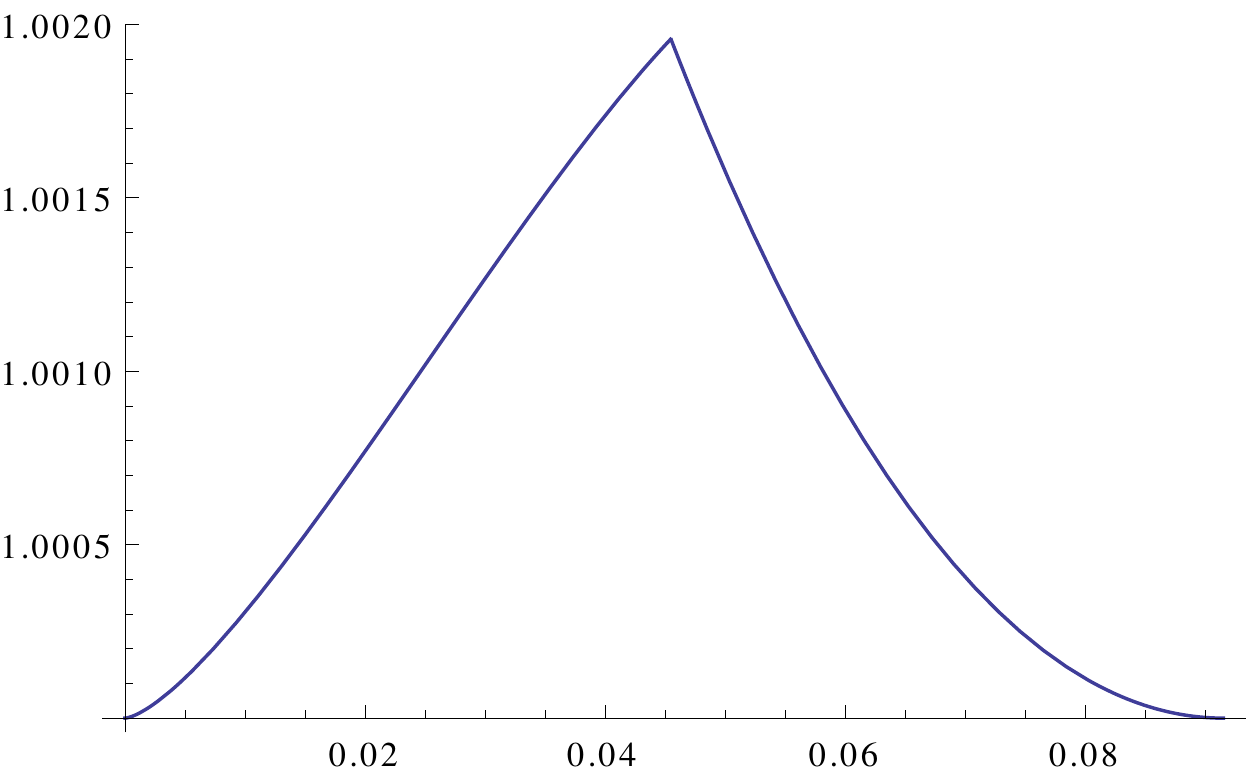}
                \caption{Best deterministic selling mechanism, $1.002$-approximate.}
                \label{fig:ratios_lp_auction_det_small_a}
        \end{subfigure}
        ~
        \begin{subfigure}[t]{0.6\textwidth}
                \includegraphics[width=\textwidth]{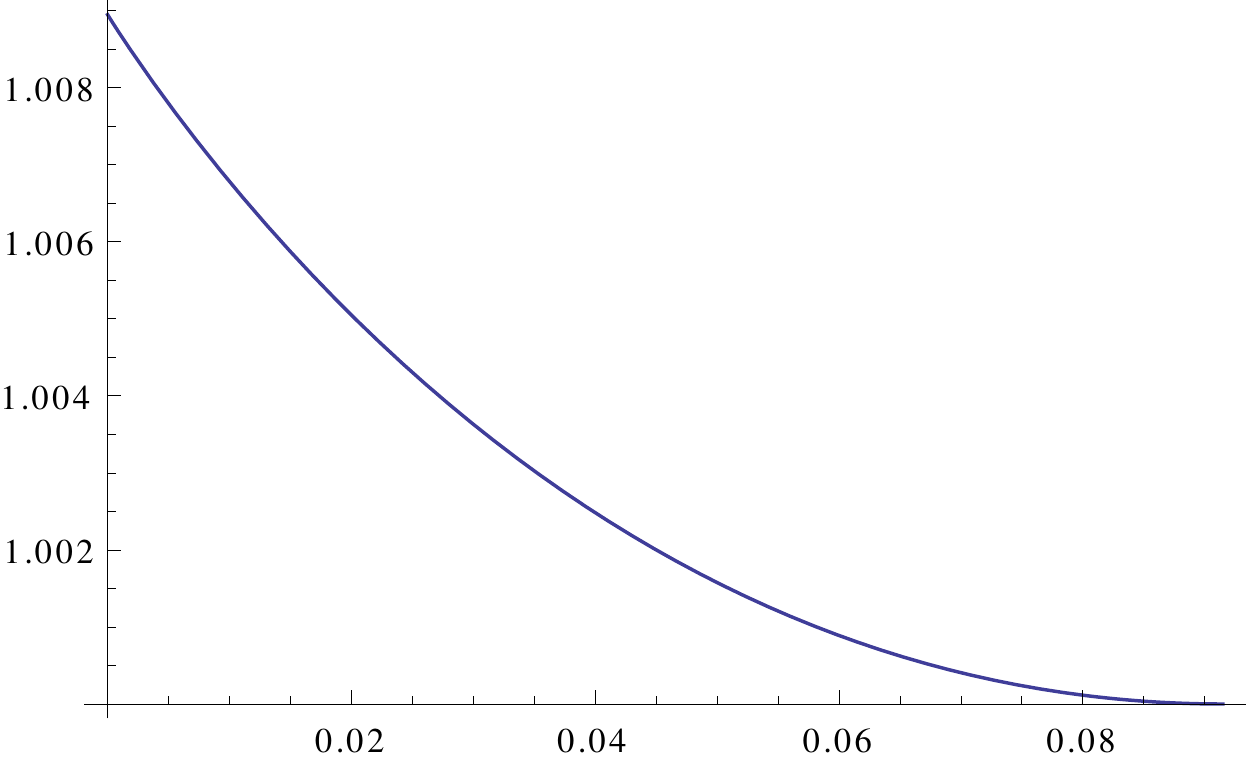}
                \caption{Best full-bundle mechanism, $1.009$-approximate.}
                \label{fig:ratios_lp_auction_bundle_small_a}
        \end{subfigure}
        \caption{Approximation ratios of the best randomized (Fig.~\ref{fig:ratios_lp_auction_rand_small_a}), deterministic (Fig.~\ref{fig:ratios_lp_auction_det_small_a}) and full-bundle (Fig.~\ref{fig:ratios_lp_auction_bundle_small_a}) selling mechanisms with respect to the optimal objective of the primal-dual approach with relaxed convexity, for $0\leq c\leq \bar c=0.092$}
        \label{fig:ratios_lp_auction_small_a}
\end{figure}

The optimality of the solutions in Fig.~\ref{fig:uniform_1} is not able to directly also give us an optimal selling mechanism, because the primal solution $\tilde u$ constructed there is \emph{not} convex. In fact, one can show that \emph{no} mechanism can achieve the primal optimal objective of $\tilde u$, which equals 
$$
\frac{2}{27} \left[\left(\sqrt{2}-4\right) c^3+3 \left(\sqrt{c (3 c+2)}+\sqrt{2}+1\right) c^2+\left(2 \sqrt{c (3 c+2)}+3 \sqrt{2}+12\right) c+\sqrt{2}+6\right]\equiv\algoname{Opt}(c)
$$
This proves that dropping the convexity constraint (in the initial formulation of our primal program) is not without loss, even in the simplest of settings: one bidder, two i.i.d.~uniform items over $[c,c+1]$ with $0\leq c\leq \bar c$. However, it turns out that this optimal objective is still not far away from the optimal mechanism's revenue, in fact it is extremely close even to that of the best deterministic or just the best full-bundle mechanism. Specifically in Fig.~\ref{fig:ratios_lp_auction_small_a} one can see that the best randomized mechanism is within a factor of 7.5\textpertenthousand, and the best deterministic and full-bundle within factors of 2\textperthousand \ and 9\textperthousand, respectively, with respect to $\algoname{Opt}(c)$.

Let's make this discussion more rigorous. As we've mentioned before, our primal-dual formulation relaxes the original optimal revenue problem in two ways: first drop the convexity constraint, that corresponds to the truthfulness requirement;
then we drop the dual variables $s_j$ that correspond to the primal constraints $\nabla u (\vecc x)\geq \vecc 0$. The latter relaxation has no actual effect to the optimal solution of the primal-dual programs, at least for the particular case we study here of i.i.d.~uniform valuations over intervals of the form $[c,c+1]$. The reason for that is simple: the optimal solution $\tilde u$ that we get after all relaxations satisfies $\nabla\tilde u(\vecc x)\geq \vecc 0$ anyways.

So now let's focus on the necessity of the crucial remaining condition, that of convexity. By \citeauthor{Pavlov:2011fk}'s results we know that for $c>0.077$ full bundling is an optimal selling mechanism. Such a mechanism in our setting sets a take-it-or-leave-it price $s$ for both items together, thus having an expected revenue of \linebreak[4] $[1-(s-2c)^2/2 ]s$ (a simple probabilistic argument, taking into consideration the area of the gray region in that case). This is maximized for $s$ being the root of $27 s^3-108 s^2 c+s \left(108 c^2-54\right)-16 c^3+4 \sqrt{2} \sqrt{\left(2 c^2+3\right)^3}+72 c=0$ giving a revenue of
$$
\algoname{BRev}(c)=\frac{2}{27} \left(-4 c^3+\sqrt{2} \sqrt{\left(2 c^2+3\right)^3}+18 c\right),
$$
which, as we said, is also the optimal revenue for $c>0.077$. However, it is easy to check that for all $0.077<c<\bar c$ it strictly holds $\algoname{BRev}(c)<\algoname{Opt}(c)$, demonstrating the gap caused by dropping convexity.

\section{The case of $c\geq 0.092$}
\label{sec:uniform_gen_2}
\begin{figure}
        \centering
        \begin{subfigure}[t]{0.6\textwidth}
                \includegraphics[width=\textwidth]{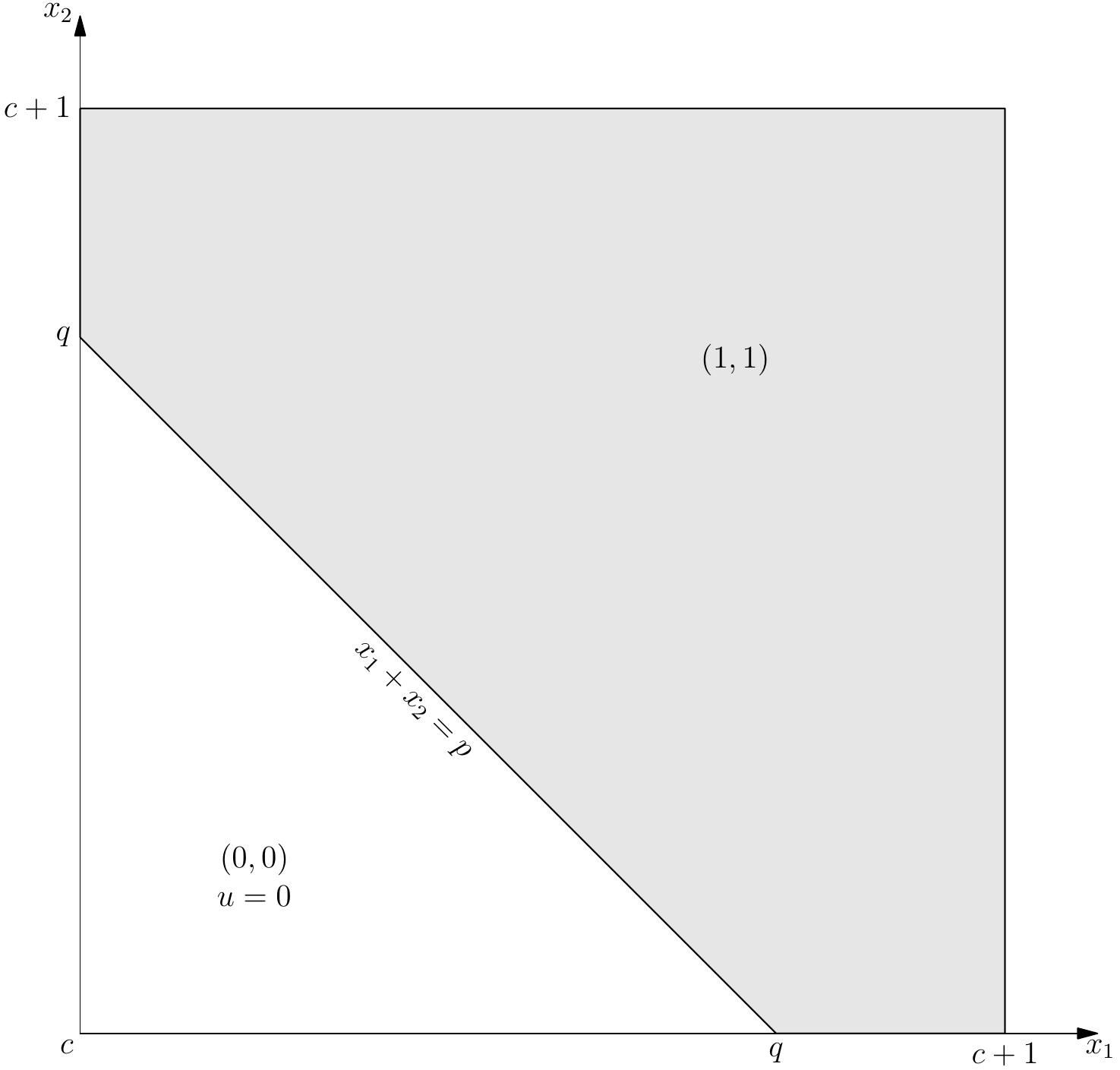}
                \caption{The values of $\nabla u(\vecc x)$ of an optimal primal solution $u$.}
                \label{fig:uniform_2_primal}
        \end{subfigure}
        ~
        \begin{subfigure}[t]{0.6\textwidth}
                \includegraphics[width=\textwidth]{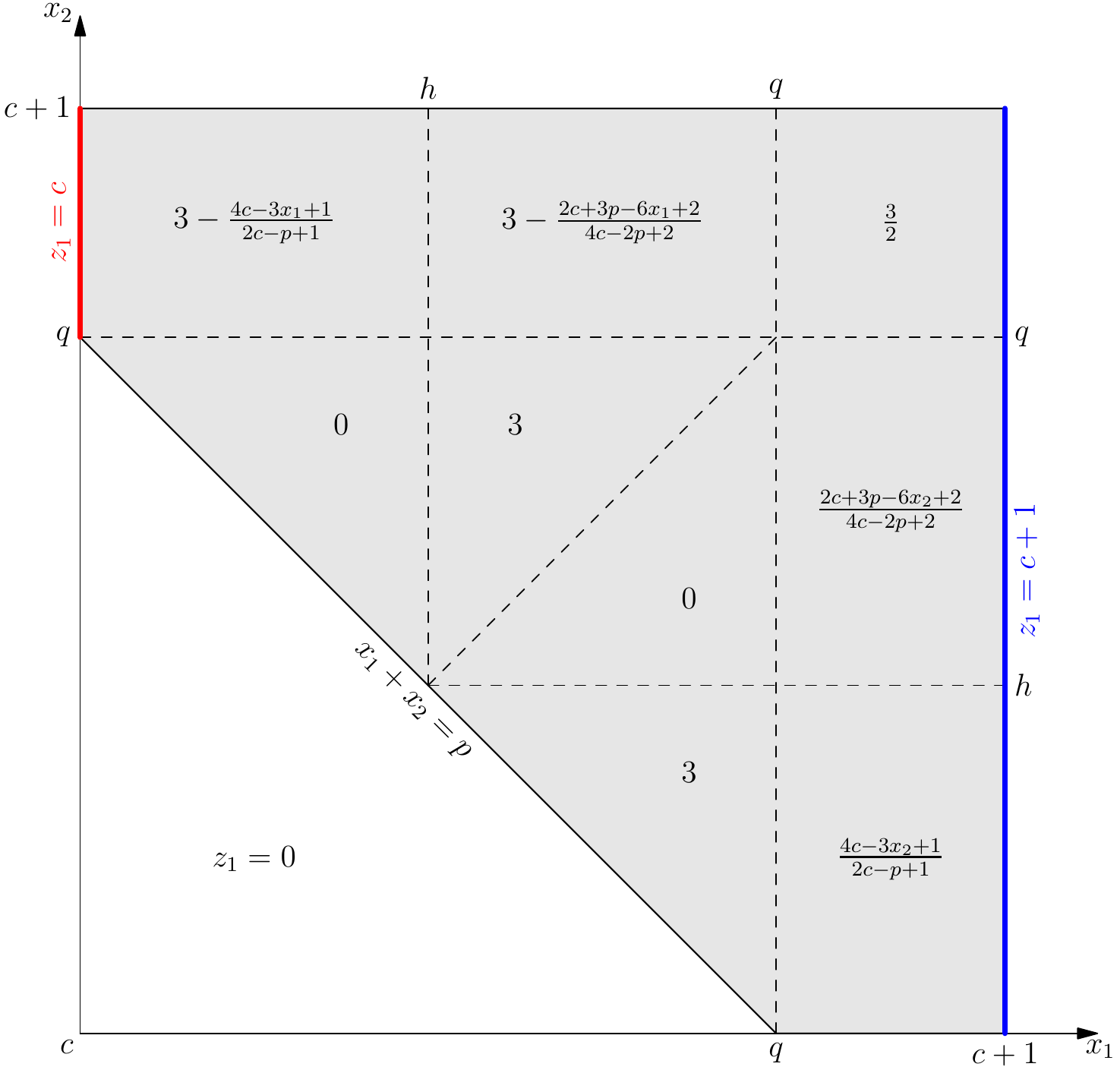}
                \caption{The values of $\partial z_1(\vecc x)/\partial x_1$. The duals are symmetric and the values of $\partial z_1(\vecc x)/\partial x_2$ can be recovered by the relation $\partial z_2/\partial x_2=3-\partial z_1/\partial x_1$ in the gray area (and $z_2=0$ in the white).}
                \label{fig:uniform_2_dual}
        \end{subfigure}
        \caption{A pair of optimal primal-dual solutions for $c\geq 0.092$. Notice the primal solution $u$ in Fig.~\ref{fig:uniform_2_primal} corresponds to a deterministic full-bundle mechanism. The values of the various parameters are: $p=\frac{1}{3} \left(4c+\sqrt{4 c^2+6}\right)$, $q=p-c$ and $h=\frac{p}{2}$.}
        \label{fig:uniform_2}
\end{figure}

For $c\geq \bar c$ it turns out that convexity is indeed not needed and the optimal solution of the primal-dual programs give also the optimal selling mechanism, which is a \emph{full-bundling} one, as can be seen by the complementarity of the pair of primal-dual solutions we propose in Fig.~\ref{fig:uniform_2}. 

It is again a matter of trivial calculations to check that indeed $z_1(c+1,x_2)=c+1$ for all $x_2\in[c,c+1]$. The point that needs more attention is making sure that $z_1$ does not get negative. There is a risk of getting below $0$ at the top stripe $q\leq x_2\leq c+1$, and in particular in the box where $a\leq x_1 \leq h$. A simple derivative argument shows us that $z_1$ achieves a minimum there at $x_1=x_1^*\equiv\frac{1}{3} \left(2 c-2\sqrt{4 c^2+6}\right)$ and solving $z_1(x_1^*,x_2)\geq 0$ we get that $c\geq \sqrt{15-8 \sqrt{2}}-2 \sqrt{2}+1=\bar c$, which is exactly the the case we are in, complementary to the previous Sect.~\ref{sec:uniform_gen_1}.  

\begin{figure}
                \centering
                \includegraphics[width=0.6\textwidth]{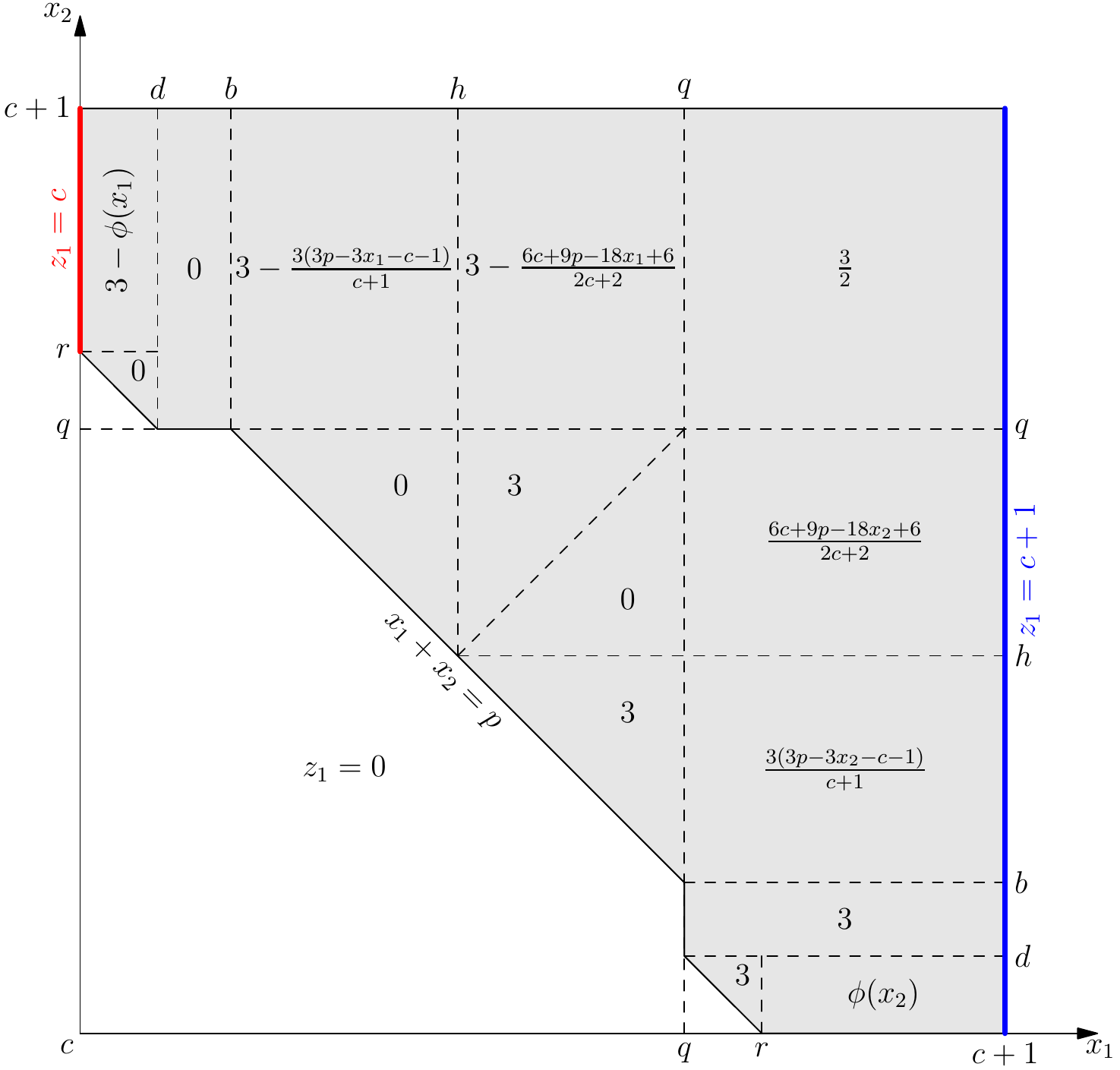}
                \caption{Alternative dual optimal solution to the one given in Fig.~\ref{fig:uniform_1_dual}. All parameters remain the same. This fits with the dual solution given in Fig.~\ref{fig:uniform_2_dual}.}
                \label{fig:uniform_1_dual_alter}
\end{figure}

We must mention here that there is also a more unified dual solution scheme that can cover both cases of Sect.~\ref{sec:uniform_gen_1} and Sect.~\ref{sec:uniform_gen_2} at the same time: simply replace the dual solution in Fig.~\ref{fig:uniform_1_dual} for $c\leq \bar c$ by the slightly more involved in Fig.~\ref{fig:uniform_1_dual_alter} which however now fits smoothly with the one in Fig.~\ref{fig:uniform_2_dual} for the other case of $c\geq \bar c$.

\paragraph{Acknowledgments} I am grateful to Anna Karlin for suggesting a triangulation of the valuation space which inspired the one presented in the current paper and simplified even more a previous dual solution of the author. I also thank Elias Koutsoupias for many fruitful discussions and guidance.

\bibliographystyle{abbrvnat} 
\bibliography{UniformTwoNote}

\end{document}